\documentclass[sigconf]{acmart}
\usepackage{booktabs} 
\usepackage{caption}
\usepackage{adjustbox}
\usepackage{subcaption}
\usepackage{relsize}
\usepackage{float}
\usepackage{url}
\usepackage{epstopdf}
\usepackage{graphics,graphicx}
\usepackage{balance}
\usepackage{array}
\usepackage{pifont}
\usepackage{multirow}
\usepackage{amsmath}
\usepackage{footmisc}
\usepackage[normalem]{ulem}
\usepackage{paralist, tabularx}

\newcommand{\new}[1]{\textcolor{blue}{#1}}

\renewcommand\footnotetextcopyrightpermission[1]{} 

\makeatletter
\def\@copyrightspace{\relax}
\makeatother

\begin{document}

\clubpenalty = 10000
\widowpenalty = 10000

\setlength{\belowdisplayskip}{0pt} 
\setlength{\belowdisplayshortskip}{0pt}
\setlength{\abovedisplayskip}{0pt} 
\setlength{\abovedisplayshortskip}{0pt}

\title[Bias in Private Label Product Recommendations on E-commerce Marketplaces]{When the Umpire is also a Player: Bias in Private Label Product Recommendations on E-commerce Marketplaces}
 
\author{Abhisek Dash}
\affiliation{
	\institution{Indian Institute of Technology Kharagpur, India}
}

\author{Abhijnan Chakraborty}
\authornote{The author was at the Max Planck Institute for Software Systems when this work was completed.}
\affiliation{
	\institution{Indian Institute of Technology Delhi, India}
}

\author{Saptarshi Ghosh}
\affiliation{
	\institution{Indian Institute of Technology Kharagpur, India}
}

\author{Animesh Mukherjee}
\affiliation{
	\institution{Indian Institute of Technology Kharagpur, India}
}

\author{Krishna P. Gummadi}
\affiliation{
	\institution{Max Planck Institute for Software Systems, Germany}
}

\renewcommand{\shortauthors}{Abhisek Dash et al.}

\begin{abstract}
Algorithmic recommendations mediate interactions between millions of customers and products (in turn, their producers and sellers) on large e-commerce marketplaces like Amazon. In recent years, the producers and sellers have raised concerns about the fairness of black-box recommendation algorithms deployed on these marketplaces. Many complaints are centered around marketplaces biasing the algorithms to preferentially favor their own {\it `private label'} products over competitors. These concerns are exacerbated as marketplaces increasingly de-emphasize or replace {\it `organic' recommendations} with ad-driven {\it `sponsored' recommendations}, which include their own private labels. While these concerns have been covered in popular press and have spawned regulatory investigations, to our knowledge, there has not been any public audit of these marketplace algorithms. In this study, we bridge this gap by performing an end-to-end systematic audit of related item recommendations on Amazon. We propose a network-centric framework to quantify and compare the biases across organic and sponsored related item recommendations. Along a number of our proposed bias measures, we find that the sponsored recommendations are significantly more biased toward Amazon private label products compared to organic recommendations. While our findings are primarily interesting to producers and sellers on Amazon, our proposed bias measures are generally useful for measuring link formation bias in any social or content networks.\footnote{\textcolor{red}{This work has been accepted at ACM FAccT 2021. Please cite the version appearing in the FAccT proceedings.}}
\end{abstract}



\maketitle
\section{Introduction}
With millions of customers relying on e-commerce platforms for their day-to-day purchase needs, many small sellers and producers depend on these platforms for their livelihood~\cite{Clement2020Retail}. 
Given the scale of these marketplaces, search and recommendation systems deployed by the e-commerce platforms effectively mediate interactions between 
customers and sellers/producers. For instance, customers often search for specific items they are aware of, land on the corresponding item pages and then follow the related item recommendations on that page to explore similar (or complimentary) items they would not have visited otherwise~\cite{pazzani2007content}. In addition to helping customers explore the item space, prior studies have shown that such recommendations are major drivers of traffic (and revenue) for many e-commerce platforms~\cite{pathak2010empirical,AmazonRINSale, sharma2015estimating}.

However, the algorithm deciding which item is related to another item is often a black box, and in recent years, different producers and sellers have raised concerns about the fairness of such black box algorithms~\cite{Aten2019Amazon,Mattioli2019Amazon,Hruska2019Amazon}. 
Platforms can explicitly design recommendations to bias the exposure of products, i.e., steer customers to items they wish to promote at the expense of other items, by linking items independent of their mutual relevance. 
This concern has been exacerbated by many e-commerce platforms (e.g., Amazon) systematically altering  (and sometimes replacing) `organic' recommendations with sponsored advertisements~\cite{Kaziukenas2019Amazon}. 
These sponsored ads are often interspersed with the organic recommendations on the product pages, making it hard to disambiguate between the two. A recent survey of more than 2,000 Amazon customers revealed that half of the respondents did not even realize being advertised to on Amazon product pages~\cite{Graham2019Amazon}. 
Thus, sponsored recommendations offer a powerful option to platforms 
to explicitly steer customers toward (or away from) certain products.

\vspace{1mm}
\noindent \textbf{How severe are the concerns? } Such concerns about biases are especially important today due to the emergence of different \textit{private label products} in e-commerce marketplaces. A private label product is produced 
and sold by the platform itself, 
providing enough monetary incentive to be discriminative against 
other products (or producers) on the platform. For example, a recent Wall Street Journal article claimed that Amazon restricts the ability of `Tier 1 Competitors' (Amazon's lingo for largest rivals in a category) 
to buy ads on their platforms when searched for private label 
gadgets like Fire TV, or Echo Show~\cite{Mattioli2020Amazon}.
E-commerce platforms' (i)~dominating control over the marketplace, which effectively makes their competing producers their customers for distribution; (ii)~vertical integration -- the fact that they both sell goods as retailers and host sales by others as a marketplace; and (iii)~ability to amass swaths of customer information, 
raises concerns of economic monopoly~\cite{faherty2017amazon, McCabe2020States, McKinnon2020Ten}. Notably, it is this last factor -- control over information -- that enhances the anti-competitive potential of the first two~\cite{khan2016amazon}. Considering the prevailing situation, a US Senator recently  remarked 
that a platform can either be an umpire or a player, but not both~\cite{Warren2020Amazon}.


These concerns are prompting policymakers 
to 
consider regulating online marketplaces~\cite{mc2019end, khan2016amazon}. For example, in a recent regulation, the Indian government has asked online marketplaces to provide {\it fair treatments} to every seller on their platforms~\cite{Govt2018FDI}. European Commission has launched an antitrust probe into Amazon's treatment of independent sellers in its platform~\cite{White2020Amazon}. Similar concerns have been raised by the US regulators. 
During the antitrust subcommittee hearings held in July 2019, Amazon was inquired whether it treats products and sellers with special relationships differently than any of their third party counterparts~\cite{Amazon2019Online}. In the subsequent report, the committee has recommended possible disentanglement of Amazon's marketplace from its private label businesses~\cite{Tracy2020House}.

\vspace{1mm}
\noindent
\textbf{Our contributions: }
Although this study can potentially be done on any e-commerce marketplace, 
here, we focus on the Amazon marketplace and study the following research question: {\it Do Amazon private label products get an unfair share of recommendations and are therefore advantaged compared to 3rd party products?}   Specifically, we consider the related item recommendations on Amazon product pages, and attempt to quantify the biases in
{\it sponsored recommendations} compared to the {\it organic (session similarity) recommendations} toward Amazon private label products.
To this end, we collect recommendation data from over 10K backpacks and 5K batteries on {\tt amazon.in},
and investigate the extent of bias towards Amazon private labels in sponsored recommendations.

We propose to model these recommendations as {\it Related Item Networks} ({\it RINs}), 
where we view item recommendations as a directed network in which nodes correspond to items, and there is a directed edge from node \textit{i} to node \textit{j}, if item \textit{j} is recommended on the page of item {\it i}. 
Note that we have two instantiations of RIN: one for organic recommendations and another for sponsored recommendations. Given the network-centric formulation of recommendations, we pose the problem of detecting bias in recommendations as follows: {\it Given two directed networks over the same set of nodes (but belonging to different groups),
how can we quantify the 
link formation bias 
towards an advantaged group of nodes (e.g., Amazon private label products in our context) in one network compared to the other}.

In this work, we investigate the link formation bias in sponsored RIN by comparing it to the organic RIN, 
assuming the latter to be an {\it unbiased baseline representation} of the real world user behavior. We propose different measures for quantifying biases: 
{\it (i) promotion bias, (ii) ranking bias, (ii) core representation bias, and (iv) exposure bias}. 
All these different bias measures consistently point to significant bias towards Amazon private label products in the sponsored recommendations.

\vspace{1mm}
\noindent
\textbf{Isn't bias in sponsored recommendations expected?} One might argue that sponsored recommendations are expected to be biased towards the products being sponsored, i.e., advertised by their corresponding sellers. However, sponsored recommendations in the context of e-commerce platforms are different from sponsored results in generic search engines for the following reasons. Sponsored and organic recommendations are far less decoupled in 
e-commerce marketplaces like Amazon. While many ads on generic online sites are \textit{complementary}, those on the e-commerce marketplaces are \textit{competitive}~\cite{Mattioli2020Amazon}.
Additionally, sponsored product recommendations 
can have significant {\it delayed impact} on the organic recommendations; 
as sales volume, click through and browsing patterns are important factors influencing the latter. Furthermore,  Amazon's private label products compete against products from other sellers and brands for ad-space on an uneven playing field, because Amazon is both the ad publisher and runs the ad exchange. 
Amazon may already have information of bids for ad-spaces from sellers of competing products as well as the ability to unilaterally reserve some ad-spaces, once again being both an 'umpire' as well as a 'player'.    
These situations may initiate a 
feedback loop that could potentially prove disadvantageous to the other 3rd party producers registered with the marketplace. Thus, we posit that the biases in sponsored results should not go unnoticed.

To the best of our knowledge, this is the first systematic audit (as opposed to the mostly speculative news articles) of the sponsored and organic recommendations on Amazon marketplace. We hope that our findings will inform and encourage further deliberations of the algorithms deployed by 
e-commerce platforms, bringing greater transparency to online marketplaces.

\section{Related Work}
\label{Sec: Related}

\textbf{Bias in recommendation systems: }
Biases in recommendations are studied in past literature~\cite{canamares2018should,chakraborty2015can,chakraborty2017who}.
For example, popular items are more likely to be recommended leading to rich get richer effects 
~\cite{canamares2018should,jannach2015recommenders}, 
and there are some attempts to mitigate the same~\cite{jannach2015recommenders,kamishima2014correcting}. 
Another 
strand of research on fairness in recommendation has emerged around the main stakeholder in recommendation framework i.e., `customers'
~\cite{yao2017beyond,zhu2018fairness,edizel2019fairecsys}. 
Recently, 
few works have advocated for fairness toward both customers and providers, 
leading to nuanced algorithms considering two-sided fairness~\cite{mehrotra2018towards,patro2020towards,geyik2019fairness,patro2020fairrec,patro2020incremental}. 

While the fairness community seems to have covered different forms of biases in recommendation frameworks, it has overlooked the special relationships that may exist between the digital marketplace and a subset of the entities, and the biases thereof. 
In this work, our goal is to audit and study the effect of the marketplace in the related item recommendation space to uncover such biases. 

\vspace{1 mm}
\noindent
\textbf{Auditing online platforms: }
Recently, researchers have begun looking at the potential societal harms posed
by opaque, algorithmic systems. 
Researchers of algorithm auditing~\cite{sandvig2014auditing} aim to produce
methodologies that enable regulators to examine black-box systems, and understand
their impact on different stakeholders. 
Mechanisms have been developed for auditing bias in ad-delivery systems~\cite{ali2019discrimination, piotr:19}, dynamic pricing~\cite{chen2016empirical}, personalization and performance of search engines~\cite{hannak2013measuring,robertson2018auditing,mehrotra2017auditing}, information segregation~\cite{chakraborty2017quantifying}, radicalization, and reachability by recommendation systems~\cite{ribeiro2020auditing, dash2019network, lamprecht2017method, dean2020recommendations} to mention a few.
However, none of the prior works have studied the biases toward entities having special relationship with the marketplace,  especially in the context of product recommendations. 

\vspace{1mm}
\noindent \textbf{Current study:} While most of the prior studies have raised concerns on different algorithms deployed on online platforms, to our knowledge, this is the first attempt to understand how perturbing related item recommendations (from organic recommendations to sponsored ad recommendations) can impact different items (and their producers) in an online marketplace like Amazon, as well as the resultant biases toward entities having special relationship with the marketplace. With respect to the algorithmic audit nomenclature / taxonomy (proposed by~\citet{sandvig2014auditing}), 
our study falls under the `\textbf{scraping audit}' category, since we rely on crawled
data to audit the effects of sponsored recommendations. Other audit methodologies are either not available
to us, or are not useful. For example, we cannot perform a
`code audit' without privileged access to Amazon recommendation 
algorithms' source codes.
\section{The Amazon Marketplace: Background and Fairness Concerns}
\label{Sec: background}

Amazon is the largest online retailer in the world,  selling thousands of products across more than 30 categories~\cite{statista2020Annual}.
There are three major stakeholders in the Amazon marketplace -- {\it customers}, {\it sellers}, and the {\it producers} of products (brands). Due to attractive loyalty programs and customer-friendly return and pricing strategies, Amazon enjoys huge brand equity among customers~\cite{Ray2015Amazon}. 
Apart from the customers, Amazon provides major business opportunities for nearly 3 million active sellers worldwide~\cite{Kaziukenas2019Amazon}. 
\subsection{Special relationships} \label{Sec: SpcRelationship}
The producers or brands 
who manufacture the products sold at the Amazon marketplace 
can be divided into two categories -- (1)~third party (3P) brands and, (2)~Amazon private label brands, and the 
corresponding products are \textbf{3P products} and {\bf Amazon private label (PL) products} respectively. 
3P brands are standalone brands (e.g., Apple, Adidas, Skybags) 
whose products are sold on Amazon under their respective brand names. 
On the other hand, Amazon private label products are produced by Amazon itself. 
Some of the Amazon-owned most successful private label brands are -- {\it AmazonBasics, Amazon Collection, Amazon Essentials, Pinzon, Solimo,} etc. 
These brands 
compete with traditional brands across multiple product categories such as electronics, fashion, jewelry etc. 
Amazon currently offers approximately $158,000$ private label products (some of which have
additional variations, such as color and size) across $45$ brands in the Amazon store~\cite{Amazon2019Online}. 
Note that as per Amazon's response to the US House antitrust subcommittee's questions, {\it all of the Amazon private label products are sold by Amazon only}~\cite{Amazon2019Online}. 
In this paper, we focus on potential biases toward such products having special relationship with Amazon.

Apart from products, some sellers also have special relationships with the Amazon marketplace, notably  \textbf{Fulfilled By Amazon (FBA)} sellers, and  \textbf{Amazon Affiliated (AA)} sellers. However, investigating the biases against sellers is confounded by other factors such as the knowledge of how sellers win a buy box\footnote{The `buy box' is shown on every product page on Amazon, and it contains the price of the product, shipping information, the name of the seller who has got the buy box, and a button to purchase the product~\cite{chen2016empirical}.} and if winning the buy box has any role to play in the recommendations received by the seller. Thus such an analysis requires its own treatment and nuances; therefore, we leave the analysis of seller bias as a future direction in this line of work.

\subsection{Concerns about potential biases}
The fact that Amazon {\it both sells (produces) goods} as a retailer, and 
{\it facilitates sales of others} as a marketplace 
allows it to amass swaths of customer information, 
giving it an edge over the competitors, thus raising concerns of economic monopoly~\cite{faherty2017amazon}. 
The primary concerns include but are not limited to (a)~PL products 
getting unfair advantage in the marketplace, and (b)~sponsored ads 
slowly replacing organic search and recommendations. which might further boost the visibility of PL products. 

\vspace{2 mm}
\noindent {\bf Concerns about PL products getting unfair advantage:}
Recent media articles, with anecdotal evidences, have argued that PL products often get promoted unduly on the Amazon marketplace through retrieval systems, which increases their exposure 
and potential sales, 
over competing products by other producers / brands~\cite{Aten2019Amazon,Mattioli2019Amazon,Hruska2019Amazon}. 

\vspace{2mm}
\noindent {\bf Concerns about sponsored recommendations replacing organic recommendations: }
Typically on Amazon, one can find various types of recommendations, including 
(i)~substitute items based on {\bf session similarity} -- ``\textit{Customers who viewed this item also viewed}'' recommendations, and 
(ii)~complementary items based on {\bf purchase similarity} -- ``\textit{Customers who bought this item also bought}''. These recommendations are `organic' recommendations, i.e., driven by customers' viewing / purchasing behavior~\cite{linden2003amazon}.  
However, in recent times, Amazon has started showing another type of recommendations -- 
(iii)~{\bf sponsored} related items--``\textit{Sponsored products related to this item}", 
which is a part of the advertising program that Amazon has taken up lately\footnote{More details on Sponsored products: \url{https://services.amazon.in/services/sponsored-products/faq.html}}. The order in which the recommendations appear is not fixed, and organic recommendations are often interspersed with sponsored advertisements~\cite{pulse2020Amazon}, making it hard to disambiguate between them.

\vspace{1mm}
\noindent
\textbf{Economic aspects of sponsored advertisements: }
One might argue that sponsored ads are meant to be biased toward promoting sellers and brands who pay for these ads. However the important point to note here is that sponsored recommendations are often non-transparent and are influenced by commercial interests. For example, it is not explicitly mentioned who pays Amazon for placing the private label ads in the sponsored recommendations, or whether Amazon allocates free ad space to its private label products. 
Upon being asked about the same in the US congress anti-trust hearing during July 2019, 
Amazon replied the following: 
``\textit{Like all retailers, Amazon makes decisions about how to use the space in its stores based on a variety of factors, centered on what customers will find most helpful. Of course, when the company chooses not to use space for advertising by third parties, Amazon foregoes the advertising fees it could have earned from that space. Deciding whether Amazon should use its store space to show ads from third parties or for merchandising placements highlighting Amazon's private brand products depends on many variables, including, for example, the customer's query, what type of product the customer is shopping for, and whether the customer is shopping on desktop, mobile, or in Amazon's app}''~\cite{Amazon2019Online}.

When private label products and other 3rd party products compete for ad-space on Amazon, it may not be a level playing field, because Amazon may already have information of the bids that others are placing for ad-spaces. Further, since Amazon is also the marketplace, it can unilaterally reserve some ad-spaces without being accountable for such reservations (as clear from the responses of Amazon). This can potentially increase the advertising cost for the remaining ad-spaces. In fact, these concerns were also raised by the US regulators wherein they asked how Amazon determines when it reserves the ad-space and what are its repercussions on the cost of advertising in remaining ad-spaces\footnote{Please refer to the questions 45-46 in~\cite{Amazon2019Online} for Amazon's official response.}. Similar concerns have also been echoed in government of India's Department for Promotion of Industry and Internal Trade (DPIIT) statement~\cite{Sengupta2020Amazon}. 

\vspace{1mm}
\noindent
\textbf{Current study: } While there are anecdotal articles regarding the unfair advantages that PL products potentially enjoy in Amazon, there is no data-driven postmortem of the platform to discover evidences (if any) of such unfair treatments. We 
perform the first systematic data-driven audit of 
both organic and sponsored recommendations for the same category of items. We show how different sponsored recommendations are from the organic recommendations through a number of measures, 
all of which seem to indicate that sponsored recommendations (ads) promote Amazon PLs  much more than the organic recommendations. 
\section{Dataset Gathered}  \label{sec:dataset}

This section discusses our strategy for collecting data from Amazon, and the dataset collected for our analyses. 

\subsection{Data collection process}\label{Sec: datacollection} 

Amazon shows various lists of related items on the webpage of every item. For the present study, we focused on collecting all recommendations shown for items in two specific categories (out of the categories defined by Amazon) -- {\bf `Backpack'} and {\bf `Battery'}. 

\vspace{1mm}
\noindent \textbf{Why these two categories?}:
We wanted to investigate categories where Amazon has several PL products 
competing with 3P products. There are many categories (as defined by Amazon itself) where there are PL products, e.g., fashion, electronics, etc. However, the range of items in these broad categories is too large, and therefore the conclusions can be misleading as well. 
Hence, we decided to analyse a category which is lower down the category hierarchy, and where Amazon is one of the largest producers in the category. 
The `Backpack' category is one such category, where Amazon produces a fair amount of products under different PL brands. 
Also `Battery' is the category where Amazon first introduced its PL products and currently holds a large fraction of the market share~\cite{MpPs2020Amazon,faherty2017amazon}.
While collecting data for a compact category like backpack / battery, we could cover most other backpacks / batteries that are in direct competition on the marketplace.

\vspace{1mm}
\noindent \textbf{Recommendations collected:} We collected the following two types of recommendations for items as shown on its webpage.
(1)~The substitutable item recommendations, that Amazon shows as  ``{\it customers who viewed this item also viewed}''. 
These are other items that are viewed by customers while looking at the source item, during the {\it same session of exploration}. Since these recommendations are obtained from natural browsing behaviour of the customer, henceforth we refer to these  as {\bf organic recommendations}. 
(2)~The recommendations shown by Amazon under the heading ``{\it Sponsored products related to this item}''. We refer to these recommendations as {\bf sponsored recommendations}.
As indicated earlier, we chose to collect these two types of recommendations specifically to understand the effects of the recently raised concern that Amazon is gradually replacing organic recommendations with sponsored ads / recommendations. 

\vspace{1mm}
\noindent \textbf{Crawling methodology:}  To collect these recommended lists, we designed crawlers performing {\it snowball sampling}~\cite{biernacki1981snowball} on the Amazon India site (\url{amazon.in}). The crawlers were seeded with an initial item, and the recommendations against the item were scraped. New items encountered in this process were pushed to a queue and the same process was repeated on every item in the queue. 
We used browser automation for our data collection. Specifically, we used Gecko driver and Mozilla Firefox browser to collect the data. All the data were collected in one session of data collection from an Amazon account having prime membership.
We continued the crawls till the queue was exhausted, to ensure that we collected the whole universe of items (or at least a large connected component of the universe of items). 
The order of visiting product pages was influenced by the order in which the products appear in the BFS queue. While visiting, the product pages, we made sure not to visit the same product page multiple times because that may have led to more recommendations for the corresponding products.
The total number of items (from each of the two categories) whose data we could collect from the Amazon website is listed in Table~\ref{netStat}. 
Along with the two types of recommendations, we also collected metadata about each product,  e.g., average user rating, number of reviews, brand, seller who won the `buy box' 
during the crawl 
etc. 

\begin{table}[t]
	\noindent
	\small
	\centering
	\begin{tabular}{ |p{1.2 cm}|p{1 cm}|p{2.3 cm}|p{2.5 cm}|}
		\hline
		{\bf Category} & \# {\bf Items} & \# {\bf Organic recos.} & \# {\bf Sponsored recos.} \\
		\hline \hline
		{\bf Backpack} & 10,775 & 375,104& 576,349\\
		\hline
		{\bf Battery} & 5,352 & 58,153 & 54,377\\
		\hline
	\end{tabular}	
	\caption{{\bf Statistics of the dataset collected from backpack and battery category products on Amazon.in.}}
	\label{netStat}
	\vspace{-10 mm}
\end{table}

\subsection{Producer/brand and related details}

There are $1593$ and $1123$ brands present in the backpack and battery datasets respectively. 
Nearly, {\bf 1.5\%  and 0.3\% of the products in the backpack and battery datasets respectively constitute Amazon PL products} which are majorly from three brands -- AmazonBasics, Amazon Brand-Solimo, and Amazon Gear. 
We collate all of them to the common brand name `Amazon'.  

In the rest of this paper, we will compare 
the `organic (session similarity) recommendations', and the `sponsored recommendations' -- for biases toward Amazon PL products due to the latter.
We took some precautions to ensure 
the meaningfulness of the comparative analysis. 
The recommendation algorithms deployed on the websites are likely to tailor 
recommendations based on several signals, including the geographical location, customer browsing history, and so on. Also the recommendations are known to vary with time. 
To minimize these variations, we collected the recommendations from the {\it same IP address} during the {\it same time-frame} 
and from the same user account.

\section{Framework for Auditing Recommendation Systems}

Motivated by the fairness concerns discussed in Section~\ref{Sec: background}, this section aims to devise methodologies based on a network-centric framework to quantify the bias toward entities having special relationships with the Amazon marketplace. 
In specific, we consider the bias toward Amazon private label products in this work; however the methodologies can be extended (if suitable data is available) to any other special relationships for e.g., relationships among different sellers and Amazon.

\subsection{The framework}

\begin{figure}[tb] 
	\begin{tabular}{|c|c|}
		\hline
		Items & Related items \\
		\hline
		A  & B, C, D \\
		\hline
		B  & E, C, F\\
		\hline
		C  & D, E, F \\
		\hline
	\end{tabular}
	\adjustimage{width=0.5cm,valign=c}{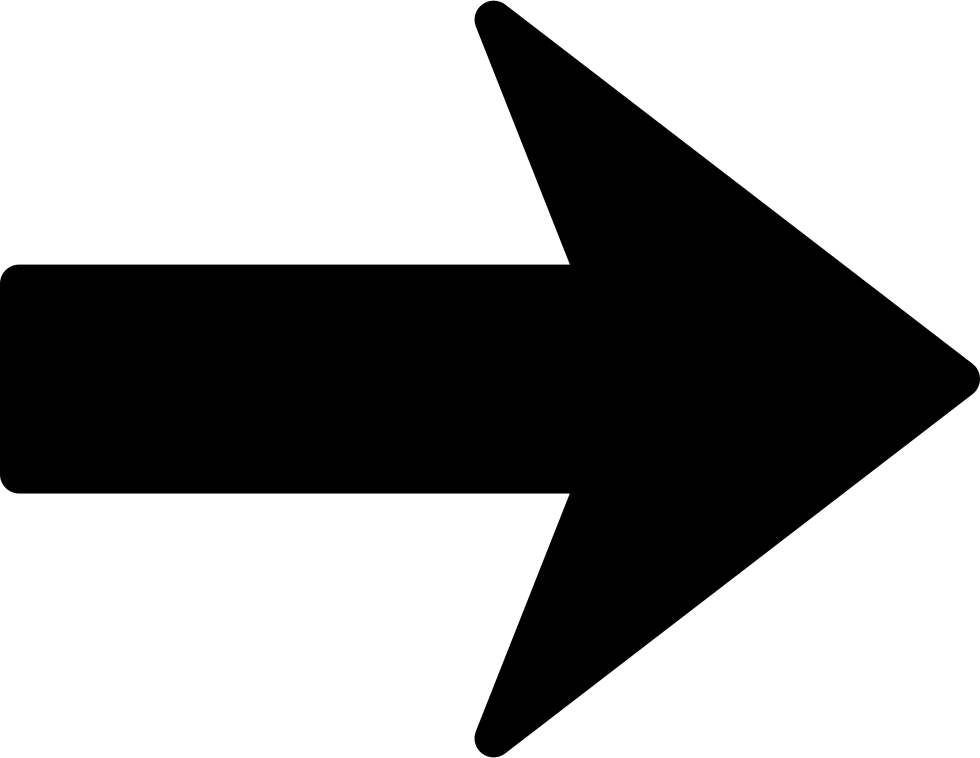}\quad
	\adjustimage{width=3.0cm,valign=c}{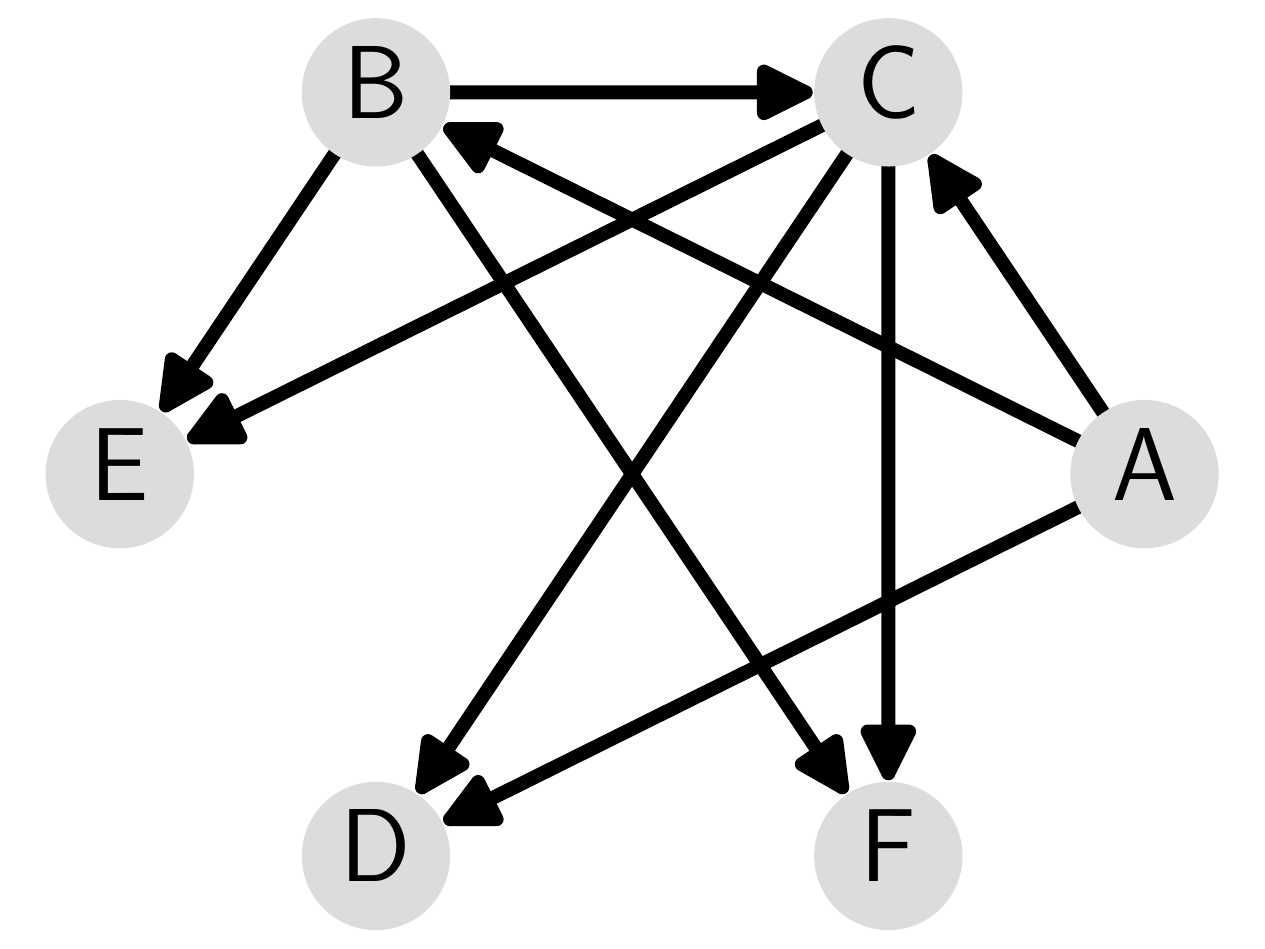}\quad
	\vspace{-2mm}
	\caption{\textbf{A sample RS showing a list of items with their related items, and its corresponding RIN.}}
	\label{Fig:RINCreation}
	\vspace{-6mm}
\end{figure}

We instantiate related item recommendations as a network~\cite{dash2019network,ribeiro2020auditing} called a {\it Related Item Network} (RIN). 
The RIN is a directed network, where each node is an item / product  
and the directed edge $i \rightarrow j$ implies that item \textit{j} is recommended on the page of item \textit{i}. 
For instance, consider the set of items and recommendations shown in the table in Figure~\ref{Fig:RINCreation}. 
These recommendations can be represented by the RIN shown in the image next to the table. 
For instance, since `B' is recommended from `A', there exists a directed edge from `A' to `B'. 
To distinguish between the RINs, we refer to the RIN constructed from sponsored recommendations as \textbf{sponsored RIN ($S$)} and that from the organic recommendations as \textbf{organic RIN ($O$)}.

\vspace{1mm}
\noindent
\textbf{Why do we require two RINs?}
Often it is difficult to quantify bias or unfairness in absolute terms. However, it is easier to establish the existence of biases when there exists an unbiased reference for comparison. As specified in Section~\ref{Sec: datacollection}, Amazon categorically labels the collected organic recommendations as items which are viewed by consumers in the same session. 
Since the organic recommendations are obtained from the most natural form of browsing behaviour of consumers, these recommendations can be thought of as a reasonable representation of the customer behavior and can be used as a baseline for further studies. However, we also acknowledge that there are several other kind of biases e.g., bias toward different types of sellers on Amazon
. As mentioned in Section~\ref{Sec: SpcRelationship}, biases toward different type of sellers is beyond the scope of the current work.
Assuming the organic (session similarity) recommendations in Amazon platform to be an appropriate and unbiased representation of user activities, we aim to quantify the relative bias (if any) toward Amazon private label products in the sponsored RIN ($S$) as compared to that in organic RIN ($O$).

\vspace{1 mm}
\noindent
\textbf{The generic research question: } Given the instantiation of recommendations as a RIN, the problem of comparing the biases in recommendations reduces to the following network-centric problem: Let $S = (S_I, S_E)$ and $O = (O_I, O_E)$ be two networks over a set of nodes, where each node has a set of attributes $A = \{a_1, a_2,..., a_m \}$. 
Without any loss of generality, let $a_s$ be a sensitive attribute 
that partitions the set of nodes (say $S_I$) into $n$ groups $G = \{g_1, g_2,...g_n \}$, where  $ \bigcup_{g_1 \ldots g_n} = S_I $ based on the membership values of the nodes (likewise for $O_I$). 
In our specific setup, the sensitive attribute is whether an item is an Amazon PL, which partitions the set of nodes into two groups (PL and 3P products).
Our aim is to quantify the relative bias due to the link formation in networks toward an advantaged group of nodes $g$ (for e.g., Amazon PLs).
Specifically in this work, we examine the relative bias induced due to sponsored RIN ($S$) with respect to the organic RIN (assuming $O$ to be an unconditional representation of the user behaviour). Since, both the sponsored and organic RINs are networks on the same set of items, henceforth we use the symbol $\mathbf{I}$ to denote the set of items i.e., for our case $S_I = O_I = \mathbf{I}$.

It can be noted that, the problem of identifying biases in networks (towards a certain group of nodes) can emerge in various other contexts as well. For example, given a pair (or more) of social networks of users, one might compare the bias due to link formation toward a socially salient group of users, in one of the networks in comparison to the other networks. The methodologies discussed in this section would be useful in such contexts as well.

\subsection{Methodologies for evaluation of bias}
In this section, we enumerate a series of methods through which we shall evaluate the relative bias (toward PL products) in the sponsored RIN with respect to the organic RIN .


\subsubsection{\textbf{Promotion bias:}} \label{Sec: PromBias}
In-degree of an item $i \in I$ in the RIN is the number of other items on whose page item $i$ is recommended. 
A higher in-degree is very likely to drive more customers toward the given item since such an item is very highly promoted on the web-page of other items. Hence we consider the promotion of a node $i$ in a recommendation system be its in-degree in the corresponding RIN. 
Specifically, promotion of node $i \in I$ in the sponsored RIN $S$ is $P_S(i) = \text{in-degree}_S(i)$, and that in the organic RIN $O$ is $P_O(i) = \text{in-degree}_O(i)$. We contrast the average $P_S(i)|_{i\in\mathbf{I}}$ with $P_O(i)|_{i\in\mathbf{I}}$ separately for the two groups of nodes (PLs vs 3Ps).

\subsubsection{\textbf{Ranking bias:} }
Network centrality measures e.g., in-degree, closeness centrality, eigenvector centrality etc., are often used to examine the most central or important nodes in networks. Having a better rank in these ranked lists is often an indication that the corresponding nodes are in some advantageous position in the current network structure. In the fairness literature, a number of measures have been proposed to quantify the bias in a given ranked list of candidates. 
We use the {\it normalized discounted KL divergence} measure ($rKL$) proposed in~\cite{yang2017measuring} to evaluate the extent of {\em mixing} of the different groups (PL and 3P products in our case) in the ranked lists. 
The  $rKL$ measure follows the notion of statistical parity fairness and evaluates whether at different check-points this statistical parity is preserved (i.e., whether PL and 3P have proportional representation at the check-points). Further, the measure also penalizes under-representation at the top-end of the ranked list more than under-representation in the lower part of the ranked list.
Given a ranked list $R$, $rKL(R) \in [0.0, 1.0]$ with $0.0$ indicating the most fair ranked list and $1.0$ indicating the most unfair ranking.

Through this analysis, we intend to contrast how good is the mixing of Amazon PL and 3P products in the ranked lists obtained from the organic RIN ($O$) and sponsored RIN ($S$). 
Let $R_s$, $R_o$ be the centrality ranked lists obtained from $S$ and $O$ respectively.
We report the ratio $\frac{rKL(R_s)}{rKL(R_o)}$. A value much higher than $1.0$ for this ratio would suggest much higher bias (i.e., the ranked list is much further away from a fair ranked list according to demographic parity) in the sponsored RIN $S$, as compared to the organic RIN $O$.

\vspace{-1 mm}
\subsubsection{\textbf{Representation in the core of the network:} }
In network science literature, nodes are often characterized by their (hierarchical) location within the network. 
Most of the highly central nodes reside in the {\it innermost core} of the network, while slightly less important nodes reside in the second innermost core, and so on, and the least central nodes reside in the {\it periphery}.
A $k$--core is a maximal sub-graph that contains nodes of degree $k$ or more. The core number of a given node $i$ is the largest value $k$ of a $k$--core containing that node~\cite{sarkar2018using}.
The nodes in the inner cores are often very strongly connected~\cite{sarkar2018rich}, and therefore a large fraction of the shortest paths connecting pairs of nodes pass through the
inner cores; the nodes in the periphery (and the outer cores) of the network are
mostly connected via the vertices residing in the innermost core of the network~\cite{sarkar2018rich,sarkar2018using}. 
Hence, {\it being in the innermost core is analogous to being visited multiple times} as users traverse the network. 

In a RIN, being in the innermost core or closer to it gives an item much more visibility as users browse the recommendations, and thereby an opportunity of larger business. To this end, if the items in the advantaged group $g$ are present in the inner cores of the network, then they enjoy the benefit of being highly exposed.
In our setup, we specifically investigate the representation of Amazon PLs in the inner (stable) cores of the organic RIN ($O$) and the sponsored RIN ($S$). To this end, any significant disparity between the representation of Amazon PLs in the inner core of the RINs can be an indication of advantageous placement of the PLs in one of the RINs. 


\vspace{-1 mm}
\subsubsection{\textbf{Exposure bias:}} \label{Sec: ExpBias} A major benefit of RS is the exposure it brings to different items among the consumers on the online system. Higher exposure is synonymous to more likelihood of conversion to sales and hence higher revenue. 
Ideally in an online system, exposure of items should be measured by the number of clicks or the total time users spend on different items. 
However, such detailed user-item interaction information is
not available to any third party (due to privacy and business concerns), making it difficult to audit/regulate  
existing recommendation systems. 
One alternative is to count the number of other items from where an item has been recommended (analogous to the in-degree of a node in the RIN) which we have already considered in the Promotion Bias methodology. 
However, it needs to be considered that all recommendations are {\it not}  of equal importance; a recommendation to item $i$ 
from a popular item is likely to give much higher exposure to $i$, than  from an unpopular item. 
Hence a model for computing exposure of $i$ should consider the importance of the items from which the recommendations to $i$ are coming. 
We use the well-known `random surfer model'~\cite{random-surfer-model} 
to incorporate such considerations.

\noindent
\textbf{Computing exposure using RIN:} 
The `random surfer model'~\cite{random-surfer-model} was originally proposed to model a user randomly surfing the Web graph, to estimate the exposure of webpages (which is measured as the PageRank~\cite{Brin98theanatomy, page1999pagerank} of a webpage). 
The random surfer model assumes that a user starts 
from a particular item, and 
then randomly chooses one of the  items recommended on this page, and views the chosen item in the next step. 
Alternatively, the user can also randomly choose any other item from 
the universe of items, for viewing in the next step. This process goes on for a large (potentially infinite) number of steps.
In the course of this random walk, some items will be visited more frequently by the surfer than other items. 
We use the random surfer model specifically because it gives a convenient way to simulate
a user exploring the item space (universe of items)
via the recommendations given to her. 
Therefore, in this methodology, we take the steady state visit frequency of a node $i$ 
as an indicator of its exposure $E(i)$ due to the RIN. 
Note that 
exposure of a node is a slight variant of its PageRank in the RIN.

We need to consider several factors to make the random surfer setting practically resemble how a user is likely to browse recommendations. For instance, users are more likely to browse higher-ranked items in a list of recommendations on the page of a certain item $i$, different users are likely to have different propensities to follow recommendations, and so on. 
We now describe these factors, and how we make the random surfer setting practical.

\vspace{1 mm}
\noindent
$\bullet$ \textbf{Effect of presentation of items on a webpage:} When a list of items is being recommended on the web-page of an item $i$, usually the {\it presentation} of these recommendations plays a vital role in the amount of exposure the recommended items get. For instance, the recommendations on the Amazon platform are shown in a slider of length $k$ (depending on the user's device).\footnote{For our experiments we set $k = 7$ as observed during the Web crawls.} 
Users have to press the `next' button to see the next $k$ recommendations. In such scenarios, the exposure of first $k$ items is 
 higher than that of the subsequent $k$ items. To account for this, we give {\it weights} to the edges of the RINs, i.e., every recommendation has a certain weight depending on the way it has been presented. To this end, based on the line of work on position bias in fairness in ranking~\cite{biega2018equity, craswell2008experimental}, we have focused on the
 	\textbf{geometric weight distribution model}, where the weights of the sliders are distributed
 	geometrically with a parameter $p$ up to the position $n$. 
 	Geometrically distributed weights are a special case of the cascade model, where each item has the same probability $p$ of being clicked\footnote{The reported results are keeping $p = 0.5$.}. This is more suitable for the current set-up because items in the first slider are more likely to be clicked than the items in the subsequent sliders.
Mathematically, if item $j$ is recommended on the page of item $i$ and item $j$ is positioned in the $t$-th slider of the recommended list on the web-page of item $i$, the weight of the directed edge $(i, j)$ is considered as $w(i,j) = p(1-p)^{t-1}, \forall t \leq n$. 

\vspace{1mm}
\noindent
$\bullet$ \textbf{Different users' propensity to follow recommendations:} \\While modeling user browsing of recommendations, one important aspect is 
the different propensity of different users to follow recommendations.
This propensity is captured in the random surfer model by the `teleportation probability' ($\alpha \in [0, 1]$). At every step, the surfer chooses one of the items recommended on the current item's page with probability $(1 - \alpha)$, and chooses to teleport to a random item in the universe with probability $\alpha$. 
We performed experiments with different values of $\alpha$ in $\{0.0, 0.1, ..., 0.5\}$
, and found the results to be qualitatively similar for all values. 


\vspace{1mm}
\noindent
$\bullet$ \textbf{Long-tail popularity distribution of items:} In the vanilla random surfer model, while teleporting 
, the random surfer jumps uniformly at random to any of the items in the network. However, this is not necessarily true in an e-commerce platform. Often people start at a popular item and then start exploring via recommendations. 
They are directed toward these popular items either via ranking in the search results on Amazon or word-of-mouth recommendations etc. To account for this phenomenon, we biased the random surfer model, so that while teleporting, instead of jumping to items uniformly at random, the random surfer follows the long-tail popularity distribution. Thereby, a popular item will have a higher chance of being teleported to. 
To operationalize the popularity of an items, we take the number of ratings (reviews) of the corresponding item as a proxy and normalize it to $[0, 1]$. 

\vspace{1mm}
\noindent
To distinguish between the exposure of an item $i$ from the two RINs, we refer to them as \textbf{Organic exposure ($E_o(i)$)} (derived from organic RIN) and \textbf{Sponsored exposure ($E_s(i)$)} (derived from sponsored RIN) respectively.
We normalize the exposure scores of all items (nodes)  in a particular network
such that, $\sum_{i \in \mathbf{I}} E_s(i) = 1$ as well as $\sum_{i \in \mathbf{I}} E_o(i) = 1$. 

\vspace{1 mm}
\noindent {\textbf{Exposure of a group of items:}}
As we defined exposure of an individual product / item, one can also compute the exposure of a group. 
For simplicity, we compute sponsored (respectively, organic) exposure of a group of items as the sum of sponsored (respectively, organic) exposure of all the products in the  group. 
The total sponsored exposure of a group $g$ can be obtained as $E_s(g) = \sum_{i \in g}{E_s(i)}$ (and similarly for organic exposure of a group). 

\noindent {\textbf{Exposure bias:}}
We define {\it Exposure Bias} ($ExpBias$) as a measure of the difference between organic exposure and sponsored exposure of items. 
We measure $ExpBias$ by the Kullback--Leibler (KL) divergence~\cite{cha2007comprehensive} between the two distributions of sponsored exposure $E_s = \{\forall i \in \mathbf{I}, E_s(i)\}$ and organic exposure $E_o = \{\forall i \in \mathbf{I}, E_o(i)\}$: 
\begin{align}
ExpBias(S) = D_{KL} (E_s||E_o)= \sum_{i\in \mathbf{I} }{E_s(i) \hspace{1mm} log \hspace{1mm} \Big(\frac{E_s(i)}{E_o(i)}\Big)}
\end{align}\normalsize
$ExpBias(S)$ quantifies the perturbation caused by sponsored recommendations on the organic exposure of different items. For a completely unbiased recommendation system, $ExpBias(S) \simeq 0$.

\vspace{2mm}
\noindent
\textbf{Categorization of items}:
Given the organic and sponsored exposures of items, we can check whether the treatment of individual items by the sponsored recommendation algorithm is {\it fair}. 
We define an item $i$ to be fairly (adequately) exposed if the ratio of its sponsored and organic exposure 
is close to $1.0$. 
Since this requirement may be too strict, we relax it by considering three item categories, depending on the ratio of sponsored and organic exposure falling within a threshold $\epsilon$ (\textbf{Relative Exposure Distortion}): \\
(a) {\bf Under-exposed:} an item $i$ \textit{under-exposed} if $1-\epsilon \leq \frac{E_s(i)}{E_o(i)}$ , \\
(b) {\bf Over-exposed:} an item $i$ \textit{over-exposed} if $\frac{E_s(i)}{E_o(i)} \geq 1+\epsilon$, and \\
(c) {\bf Adequately-exposed:} If $E_s(i)$ is between $1-\epsilon$ and $1+\epsilon$ of $E_o(i)$, then item $i$ is considered to be \textit{adequately exposed}.

\vspace{1mm} \noindent
In this paper, we use $\epsilon=0.2$ (20\%) for proof of concept. We believe any slight variation to $\epsilon$ will not change our observations significantly. 
While this 
threshold has been used in multiple prior studies~\cite{chakraborty2017who, dash2018beyond}, we acknowledge that the choice remains context dependent and can change based on the application and prior established regulations.
\subsection{Influence of the sensitive attribute} \label{sub:nbr-method}

Till now in this section, we have proposed a number of methodologies to evaluate the relative bias toward the advantaged group ($g$) in the sponsored RIN ($S$) as compared to the organic RIN ($O$). However, we have not yet touched upon the influence or effect (if any) of the sensitive attribute ($a_s$) on the treatment that different groups of items receive in the two RINs. 
Now, we describe a methodology to quantify how much influence the sensitive attribute (e.g., Amazon private label membership) has on the beneficial treatments of the products (if any).
Note that reception of any benefit (e.g., better promotion, or rank, or core number, or exposure) by a group of products might not always be an indication of bias. Let us consider a scenario where the items, which are of higher quality, are being advantaged more in the sponsored networks. In that case, the platform (Amazon in our case) might argue that since these items are qualitatively better, therefore to improve consumer experience they are being promoted excessively. 
Thus, along with quantifying biases / benefits in exposure received by certain items, it is equally important to investigate which attributes of the items have the most influence on the benefits. To this end, the following experiment is meant for explaining the importance of the sensitive attribute ($a_s$) rather than quantifying any biases based on it.

\vspace{1 mm}
\noindent
\textbf{Intuition behind the analysis:} As mentioned earlier, each node $i$ has a set of attributes $A(i)=\{a_1(i), a_2(i),..., a_m(i)\}$. Without any loss of generality, let us assume $a_s$ is a sensitive attribute e.g., gender, or race, or in our context existence of the special relation with the Amazon marketplace. 
The aim of this analysis is to find out the difference in the relative importance of different attributes $a_t(i), \forall t \in \{1..s\}$ on any beneficial commodity (e.g., better promotion or exposure of nodes) between the organic and sponsored RINs. 
Through this analysis, if we find a significant rise in the importance of the sensitive attribute $a_s$ in determining the benefit of a node, 
this can be an indication of some special treatments toward a group (on the basis of $a_s$) during the link formation in the sponsored RIN ($S$) as compared to the organic RIN ($O$).
To this end, we use Negative Binomial Regression (NBR)~\cite{hilbe2011negative} to perform this analysis. 

\vspace{1 mm}
\noindent
\textbf{NBR based approach:} NBR is a generalization of Poisson regression which slacks the restrictive assumption that the variance is equal to the mean made by the later~\cite{hilbe2011negative}. Given a count-type dependent variable $P_s(i)$ (or $P_O(i)$, and a set of attributes ($A(i)$), the model predicts which attributes correlate more with the dependent variable.
For each attribute, the model gives a numeric estimate (can be positive as well as negative) of how strongly it correlates with the dependent variable. A mathematical interpretation of these results can be, by 1 unit change in an attribute (say, $a_1(i)$), the log of $P_S(i)$ (or $P_O(i)$) will increase (or decrease) by the estimated number of units.
If the sensitive attribute is given a high numeric estimate, relative to other attributes, then the sensitive attribute can be said to have a selectively high influence on the values of $P_S(i)$ (or $P_O(i)$).

\section{Results}
In this section, we apply the proposed methodologies on the collected datasets 
to examine the relative bias in the sponsored RIN ($S$) with respect to the organic RIN ($O$) toward Amazon PL products.

\subsection{Promotion bias}\label{Sec: ResultPromBias} 
In our first bias measure, we investigate the promotion bias in the sponsored RINs toward Amazon PLs. Note that promotion of an item in the RIN is its in-degree in the corresponding RIN. A higher in-degree is likely to drive more customers toward the given item.

\vspace{1 mm}
\noindent 
\textbf{Not all 3P products are promoted in Sponsored RIN:} Intuitively, we would expect a much smaller fraction of
products to be sponsored (advertised) on a platform. 
Indeed, we observe that a very small fraction of the entire number of products that we collected have non-zero in-degree in the sponsored RINs i.e., were advertised at least once. 
Specifically, more than 50\% of the 3P products do {\it not} even get a single inward recommendation in the sponsored RINs in both backpack and battery categories. 
However, almost all Amazon PLs \textbf{($\ge 94\%$)} are recommended in the sponsored RIN in both the categories.
Thus Amazon does significant self-promotion for almost all their PL products in the sponsored RIN. 

\vspace{1 mm}
\noindent
\textbf{Amazon PLs get sponsored recommendation from half of the product space:}
We observe that Amazon PLs get recommended from approximately 50\% distinct products in the entire product base in both categories (in contrast to approximately 15\% in organic recommendations). In other words, almost half of the entire product space recommends at least one Amazon PL in its sponsored recommendations. Whereas, a very small fraction of the entire product space is recommended back from the Amazon PLs (less than 15\% in both the categories) in the sponsored recommendations.

\begin{figure}[t]
	\centering
	\begin{subfigure}{0.5\columnwidth}
		\includegraphics[width= \textwidth, height=3cm]{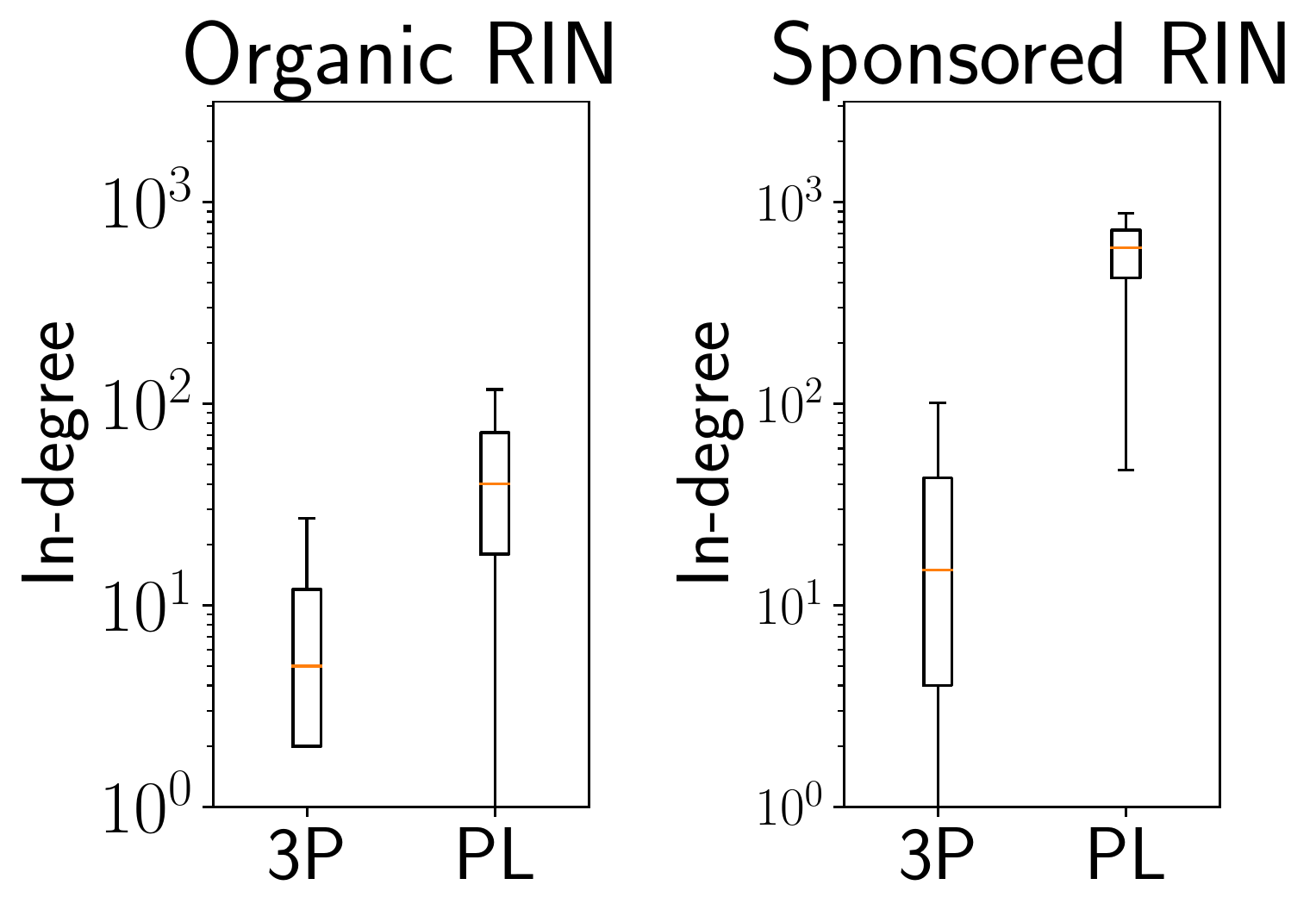}
		\vspace*{-4mm}
		\caption{}
		\label{Fig: Organic}
	\end{subfigure}%
	~\begin{subfigure}{0.5\columnwidth}
		\centering
		\includegraphics[width= \textwidth, height=3cm]{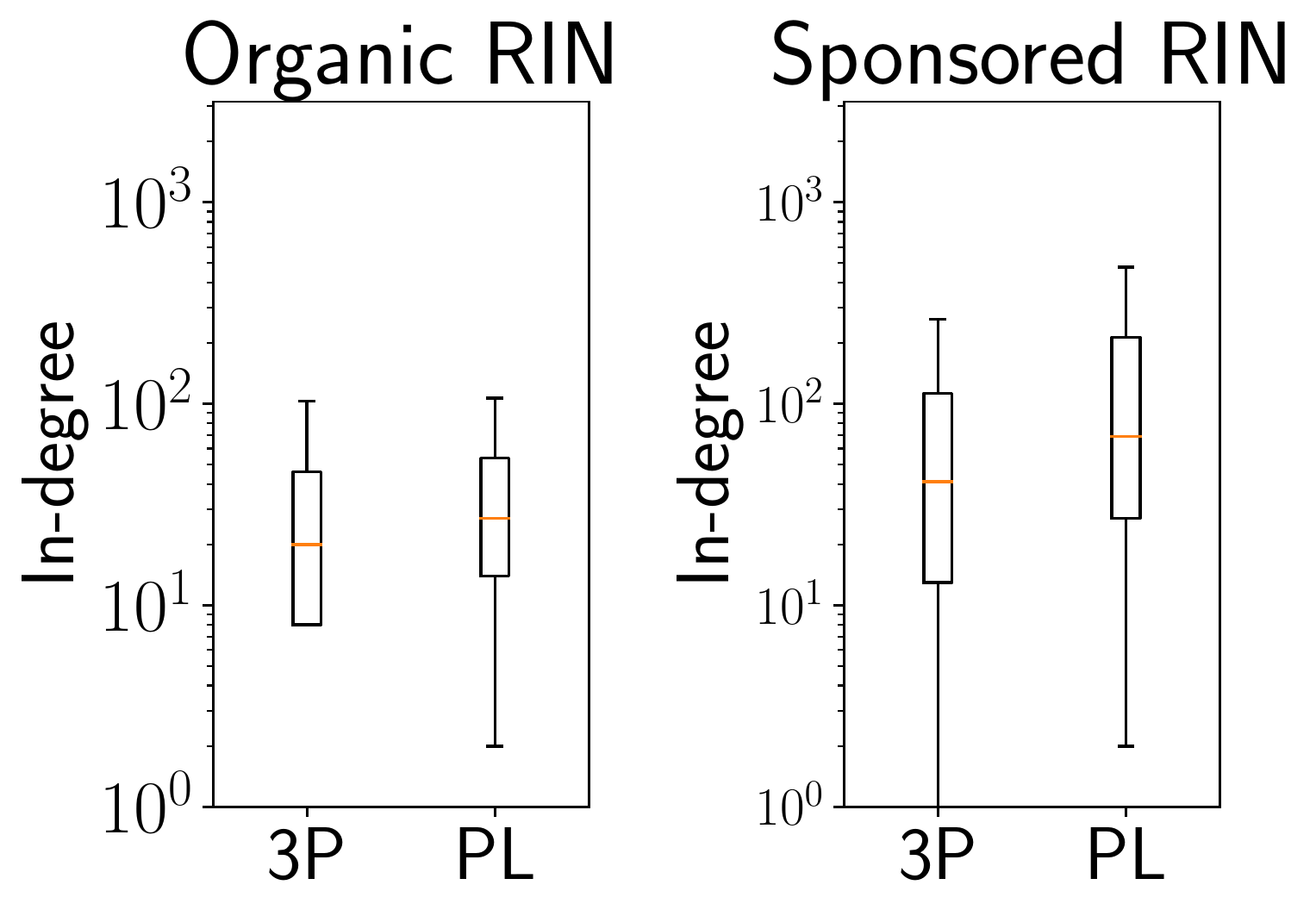}
		\caption{}
		\label{Fig: Sponsored}
	\end{subfigure}
	\vspace*{-4mm}
	\caption{{\bf Box-plots showing the median, first and third quartile of in-degree distribution of Amazon PL products and that of 3P products in (a)~battery and (b)~backpack categroy RINs. Note that the $y$-axes are in log scale. The disparity in the in-degree distributions of PLs and 3P products increase significantly in the sponsored RIN across both categories.}}
	\label{Fig: DegDist}
\end{figure}

\noindent
\textbf{In-degrees of different types of products:} Figure~\ref{Fig: DegDist} shows box-plots of in-degree distributions of all the nodes having non-zero in-degree. All nodes (items) have been divided into two groups -- Amazon PLs and  3P  products. 
In the battery category, there is a considerable disparity in the in-degree distributions of PL and 3P products in the organic RIN; however the disparity increases significantly in case of the sponsored RIN.
Interestingly, in the organic RIN of backpack category, the in-degree distributions of both 3P and PL products are very similar; however when we move to sponsored RIN there is distinguishable difference in the in-degree distributions. 
These box-plots clearly show the rise in disparity of the in-degree or promotion of PL products in the sponsored RIN as compared to the organic RIN. 
Moreover, the difference between the in-degree distributions of 3P and PL products were found to be statistically significant in the sponsored RIN according to students' t-test ($p \simeq 0.0$ for battery and $p\le0.005$ for backpack).

Further, to explore the promotion bias toward Amazon PL products in sponsored RIN, we look into the average in-degree of different types of products based on their special relationship with the marketplace. 
Table~\ref{Tab: AvgIndeg} lists the average in-degree of different types of products in the RINs. 
\begin{table}[t]
	\noindent
	\small
	\centering
	\begin{tabular}{ |p{1.9 cm}|p{1.25 cm}|p{1.25 cm}||p{1.25 cm}|p{1.25 cm}|}
		\hline
		\textbf{Avg. in-degree} & \multicolumn{2}{c||}{\textbf{Backpack}}  & \multicolumn{2}{c|}{\textbf{Battery}}\\
		\hline
		& Organic & Sponsored & Organic & Sponsored \\ 
		\hline		
		All nodes & 35 & 54 & 11 & 11 \\
		\hline \hline
		\multicolumn{5}{|c|}{\textbf{Product category}} \\
		\hline \hline 
		Amazon PLs & 45 & 164 & 46 & 520 \\
		\hline
		3P products & 34 & 52 & 11 & 09 \\
		\hline  
	\end{tabular}	
	\caption{{\bf Average in-degree for different kinds of products across sponsored and
			organic RINs. The average in-degree of PL products sees tremendous increase in the sponsored RIN. The differences in the in-degrees were statistically significant (students' t-test).}}
	\label{Tab: AvgIndeg}
	\vspace{-10 mm}
\end{table}
For the Battery category, an Amazon PL is recommended from 46 other items on average in the Organic RIN; however in the Sponsored RIN of the same category, an Amazon PL is recommended from as many as $520$ other items. In sharp contrast, a 3P battery product gets recommended from only $9$ other items in the Sponsored RIN on average, which is actually {\it lower} than the  average number of $11$ items recommending a 3P battery in the Organic RIN.
Similarly, for the Backpack category, Amazon PLs receive recommendations from many more items on average  in the Sponsored RIN ($164$) as compared to the Organic RIN ($45$). 

These values clearly suggest that the increase in the degree of promotion of Amazon PL products is significantly higher in Sponsored RINS, than that of the 3P products.
In other words, products that have special relationships with Amazon receive disproportionately more sponsored recommendations than organic recommendations. 


\subsection{Ranking bias }\label{Sec: ResultRankBias}
To quantify the ranking bias of the RINs, we require a centrality measure of choice and a sensitive attribute. To this end, we performed our analyses across different choices of network centrality measures. For brevity, we explain the results for in-degree centrality~\cite{freeman1978centrality} measure and sensitive attribute based on the PL membership of the products i.e., whether a product is an Amazon private label (PL) or a 3P product. 

\noindent
\textbf{In-degree centrality:} In-degree centrality~\cite{freeman1978centrality} of a node in a network is the fraction of inward edges coming to it. Hence, a better rank in this centrality measure suggests that a particular node can be reached with higher likelihood from the rest of the network. In the context of RINs, better in-degree centrality is synonymous to higher likelihood of customer's visibility for products.

\noindent
\textbf{Observations: }
Amazon PLs get significantly better centrality ranks in the sponsored RINs than in the organic RIN. 
In both categories (battery and backpack), the ranked list obtained from the sponsored RIN i.e. $R_s$ is almost $8$ times as biased as the ranked list from the organic RIN ($R_o$). 
Specifically, in terms of the normalized discounted KL divergence measure explained in the previous section, $\frac{rKL(R_s)}{rKL(R_o)} = 7.8$ and $7.6$ respectively for the battery and backpack categories. 
Also, in both the battery and backpack categories,
more than 80\% of all Amazon PL products have higher rankings in the Sponsored RIN, than their ranking in the Organic RIN.
Thus, the ranked lists obtained from the sponsored RIN ($R_s$) shows significant bias by placing Amazon PL products in better ranks.

\subsection{Representation in the core of the network }\label{Sec: ResultCore}

In this methodology, we calculate the core number of the PL and 3P products. 
Note that higher core-numbers indicate that the corresponding nodes are located in the inner cores, and are hence more strategically placed to derive higher exposure. 
We observe that the PL products 
heavily occupy the inner cores in the sponsored RIN. 
For the backpack category, there are as many as 12\% of all Amazon PLs in the innermost  core (core number = 85) of the sponsored RIN; in contrast, there was no Amazon PL in the innermost core of the organic backpack RIN. 
In case of battery category, both the RINs contained 17.65\% of all Amazon PLs in their respective innermost core. Further, if we divide the cores into four quartiles, we observe that in the sponsored RIN, 94\% of PLs in battery category (67\% in backpack category) are in the top 2 quartiles (i.e. $coreNumber(i) \ge 0.5 * max(coreNumbers)$). The percentage reduces to mere 65\% in the organic RIN (57\% in backpack category).

\begin{figure}[t]
	\centering
	\begin{subfigure}{0.48\columnwidth}
		\includegraphics[width= \textwidth, height=3cm]{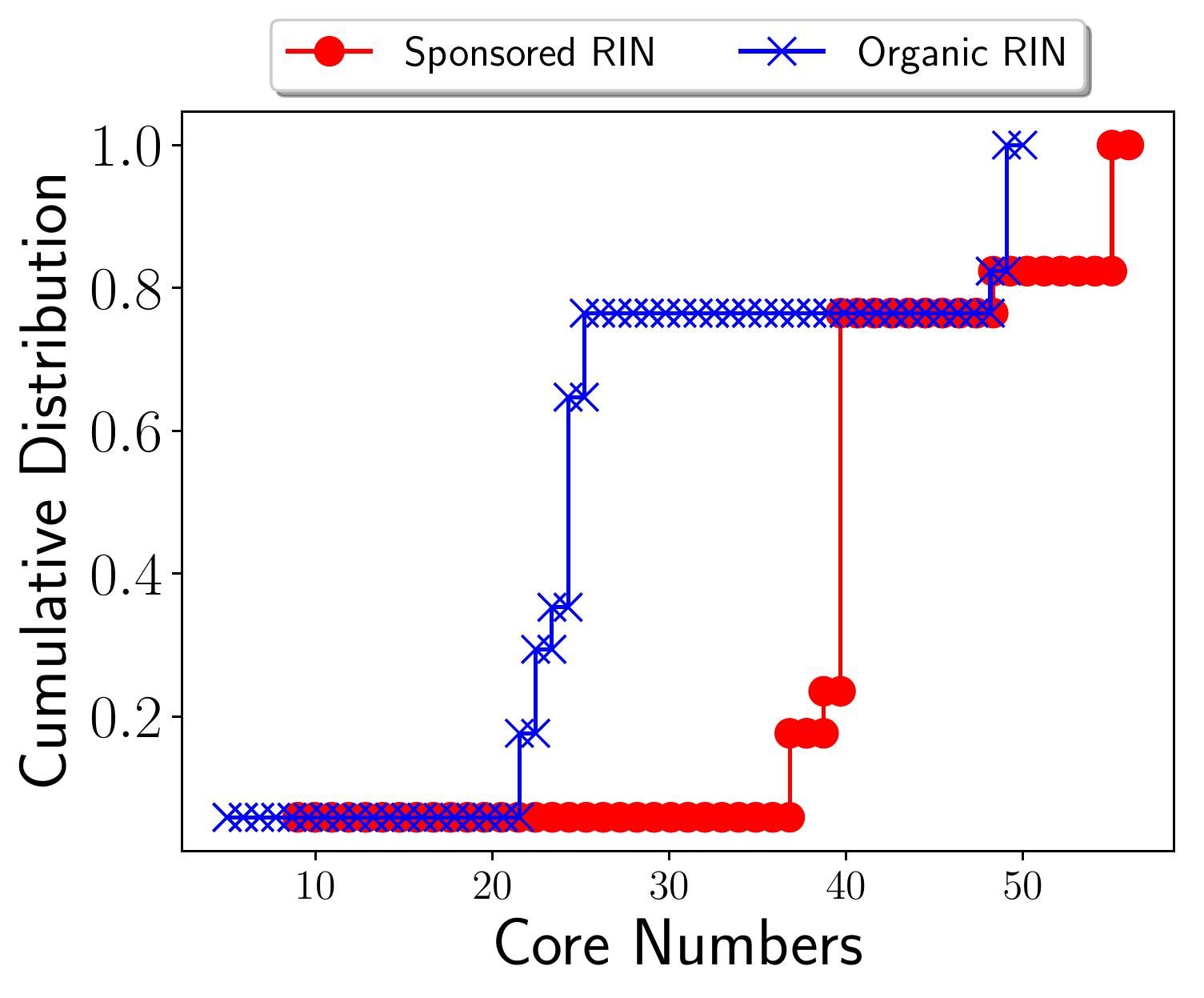}
		\vspace*{-4mm}
		\caption{}
		\label{Fig: CorePLBattery}
	\end{subfigure}%
	~\begin{subfigure}{0.48\columnwidth}
		\centering
		\includegraphics[width= \textwidth, height=3cm]{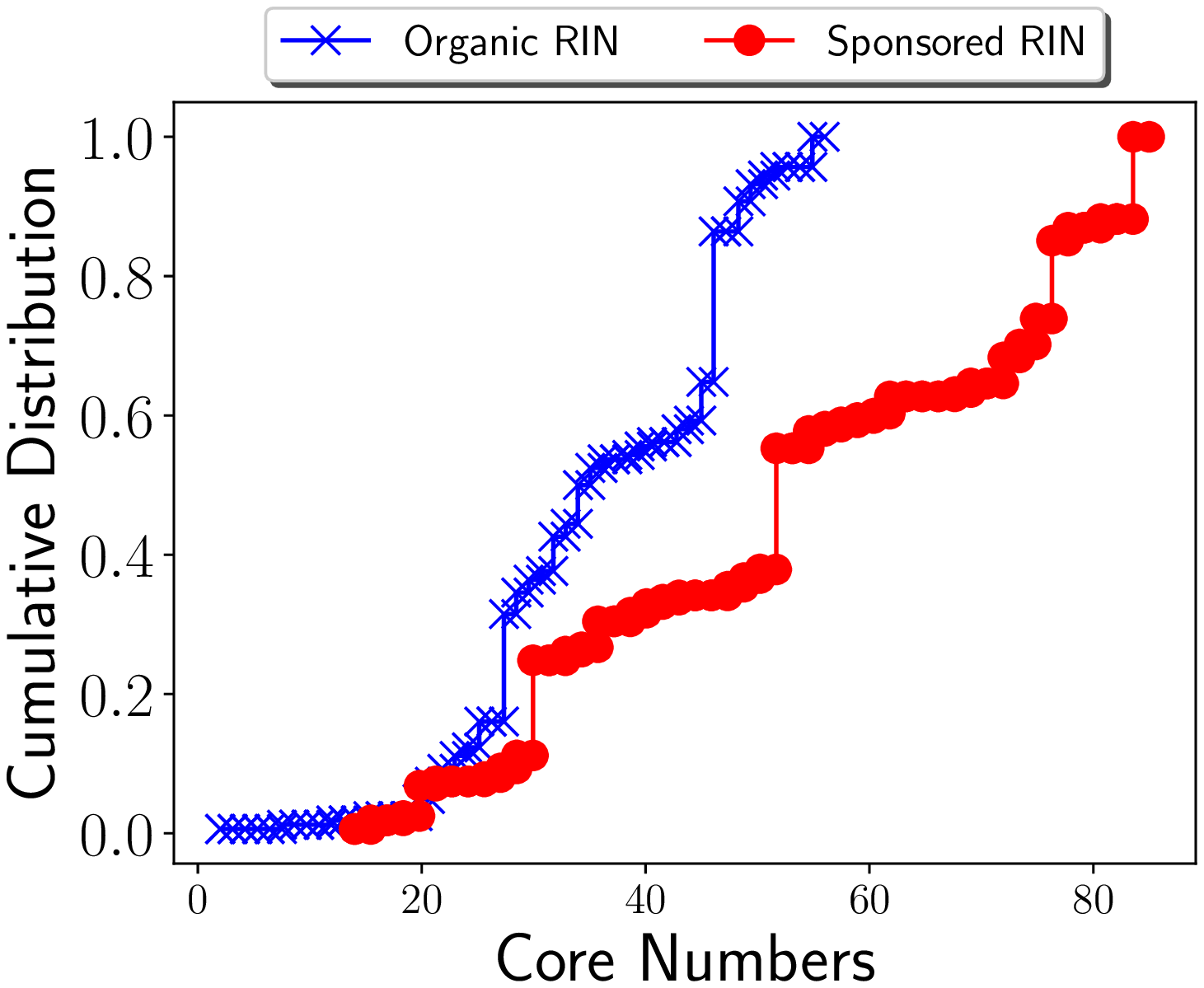}
		\vspace*{-4mm}
		\caption{}
		\label{Fig: CorePLBackpack}
	\end{subfigure}
	\vspace*{-2mm}
	\caption{{\bf ECDF of the core numbers of PL products in (a)~battery and (b)~backpack category. In general, PL products are placed closer to the innermost core in the sponsored RINs.}}
	\label{Fig: CorePeriphery}
	\vspace{-3mm}
	
\end{figure}

Figure~\ref{Fig: CorePeriphery} shows the cumulative distribution of core numbers of PLs in both RINs of the (a)~battery and (b)~backpack categories. The figures show that the Amazon PL products are 
placed closer to the innermost core in the sponsored RIN to increase their overall visibility, as compared to the organic RIN.

\subsection{Exposure bias}\label{Sec: ResultExpBias}

The exposure bias scores ($ExpBias$) are computed using the customised Random Surfer model, as described in  Section~\ref{Sec: ExpBias}. 
First, we report the results by varying the teleportation probability $\alpha$ in $\{0.0, 0.1, ..., 0.5\}$ of the random surfer model to evaluate the exposure of items in the corresponding RINs. 

The primary observations from Table~\ref{Tab: AlphaVariation} is that with the increase in teleportation probability ($\alpha$), percentage of nodes adequately exposed increases gradually and that of under exposed nodes decreases. This is primarily because, the percentage of times the random surfer is allowed to teleport (sample a random node) increases with the increase in $\alpha$.
However, a common trend that remains the same is that the majority of the nodes still remain under-exposed in the sponsored RIN as compared to the organic RIN. 
Throughout the rest of this paper, we report results by keeping $\alpha = 0.15$ (which is the default value prescribed in the random surfer model~\cite{page1999pagerank}). 

\begin{table}[tb]
	\noindent
	\small
	\centering
	\begin{tabular}{ |p {0.5 cm}|p {1.6 cm}| p {2.25 cm}|p {1.75 cm}|p{1 cm}|}
		\hline 
		\bf $\alpha$ & Over exposed &Adequately exposed & Under exposed & ExpBias\\
		\hline \hline
		\multicolumn{5}{|c|}{\textbf{Battery category}} \\
		\hline \hline
		0.0 & 14.9 & 2.3 & 82.8 & 2.76\\
		\hline
		0.1 & 12.6 & 8.9 & 78.5 & 2.29\\
		\hline
		0.2 & 13.2 & 10.6 & 76.2 & 1.72\\
		\hline
		0.3 & 13.5 & 12.5 & 74.0 & 1.36\\
		\hline
		0.4 & 14.0 & 15.0 & 71.0 & 1.08\\
		\hline
		0.5 & 13.9 & 18.6 & 67.5 & 0.85\\
		\hline \hline
		\multicolumn{5}{|c|}{\textbf{Backpack category}} \\
		\hline \hline
		0.0 & 20.6 & 3.4 & 76.0 & 2.46\\
		\hline
		0.1 & 16.4 & 14.6 & 69.0 & 1.64\\
		\hline
		0.2 & 17.3 & 17.6 & 65.1 & 1.26\\
		\hline
		0.3 & 17.8 & 21.0 & 61.2 & 0.99\\
		\hline
		0.4 & 18.2 & 24.2 & 57.6 & 0.78\\
		\hline
		0.5 & 18.2 & 28.3 & 53.5 & 0.53\\
		\hline
	\end{tabular}	
	\caption{{\bf Percentage of items over, adequately and under-exposed in the sponsored RIN along with the induced exposure bias in comparison with the organic RIN with different teleportation probabilities ($\alpha$). The results are qualitatively similar for all $\alpha$ values i.e., majority of the products are being under-exposed in sponsored RIN.}}
	\label{Tab: AlphaVariation}
	\vspace{-6mm}
\end{table}

The exposure bias scores for both the categories  (battery and backpack) with $\alpha = 0.15$ are listed in Table~\ref{Tab: expStat}. 
Sponsored recommendations in battery category induce more exposure bias than backpack category. We observe that almost 68\% items in the backpack category get significantly under-exposed and the percentage rises to 76\% in the battery category. 
The overall consequences that such distortions can have on the producers of different items is also worth investigating.

\begin{table}[t]
	\noindent
	\small
	\centering
	\begin{tabular}{ |p{2.3 cm}|p{1.2 cm}|p{1.2 cm}||p{1.2 cm}|p{1.2 cm}|}
		\hline
		\textbf{Exp. categories} & \multicolumn{2}{c||}{\textbf{\% of items}}  & \multicolumn{2}{c|}{\textbf{ExpBias}}\\
		\hline
		& Battery & Backpack & Battery & Backpack \\ 
		\hline		
		Over Exposed & 12.72\% & 18.39\% & & \\
		\cline{1-3}
		Adequately Exposed & 10.81\% & 13.39\% & 2.19 & 1.27\\
		\cline{1-3}
		Under Exposed & 76.47\% & 68.22\% & & \\
		\hline
	\end{tabular}	
	\caption{{\bf Percentage of items over, adequately and under-exposed in the sponsored RIN along with their induced exposure bias in comparison to the organic RIN (with $\alpha = 0.15$).}}
	\label{Tab: expStat}
	\vspace{-5 mm}
\end{table}

\begin{figure}[t]
	\centering
	\begin{subfigure}{0.48\columnwidth}
		\includegraphics[width= \textwidth, height=4cm]{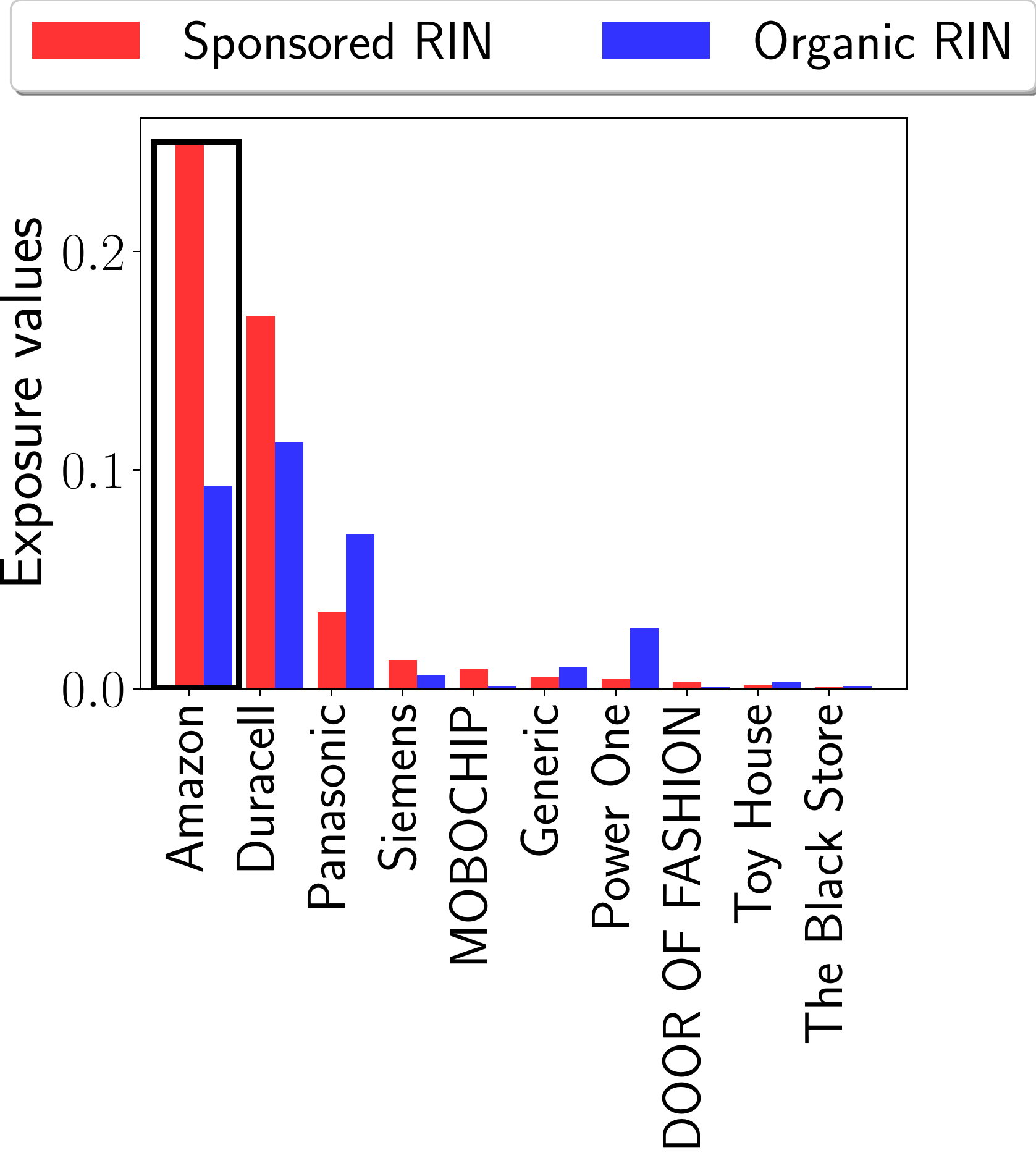}
		\vspace*{-4mm}
		\caption{}
		\label{Fig: AbsBrandBattery}
	\end{subfigure}%
	\hfill
	~\begin{subfigure}{0.48\columnwidth}
		\centering
		\includegraphics[width= \textwidth, height=4cm]{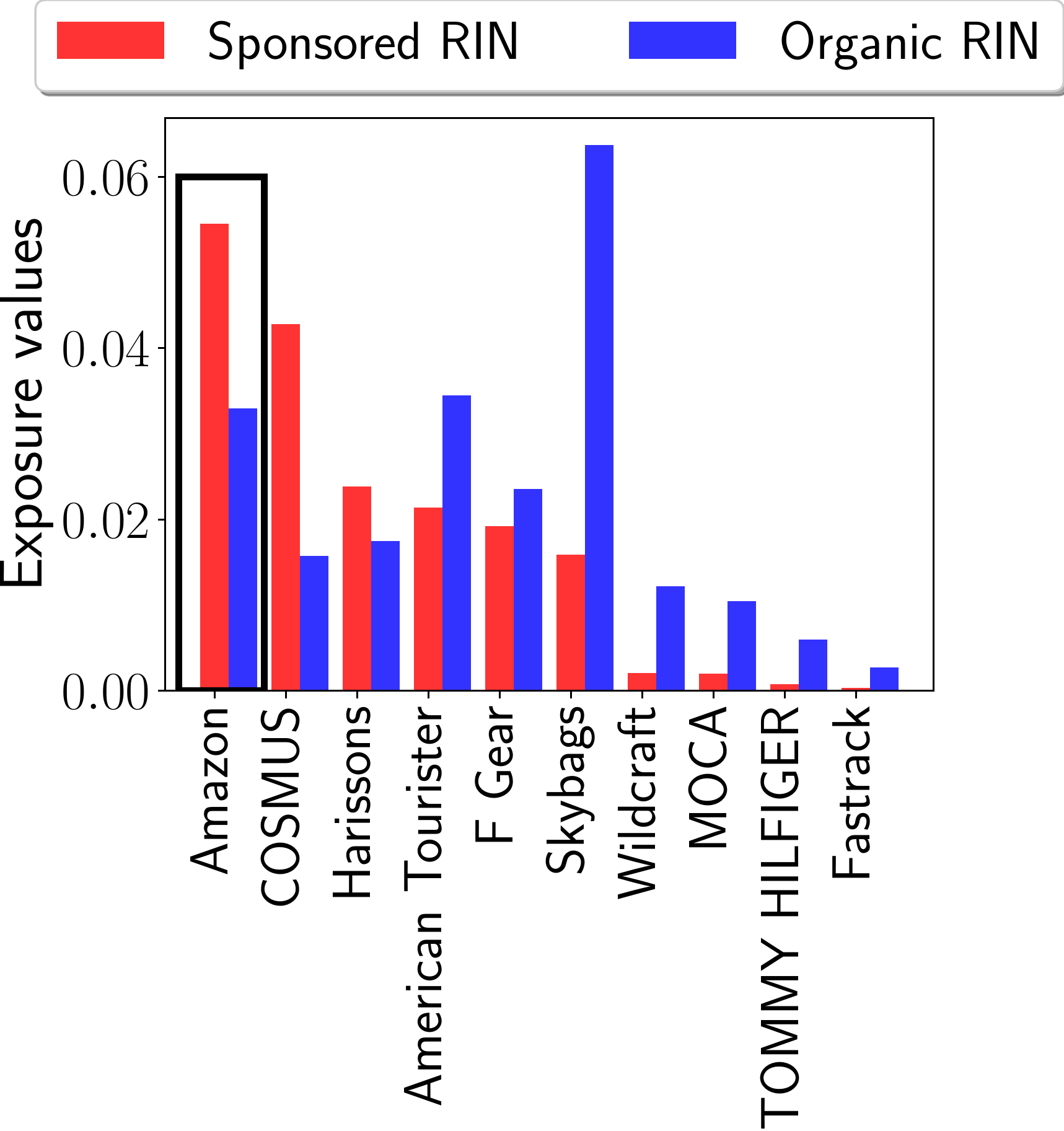}
		\vspace*{-4mm}
		\caption{}
		\label{Fig: AbsBrandBackpack}
	\end{subfigure}
	\vspace*{-2mm}
	\caption{{\bf Top 10 brands in (a)~battery, and (b)~backpack categories and their absolute exposure in sponsored and organic RINs. Amazon private label brands (highlighted with a rectangle) account for nearly 25\% and 5.5\% of the total exposure in sponsored RINs in the battery and backpack categories.}}
	\label{Fig: AbsExpBrand}
	\vspace{-3 mm}
\end{figure}

\vspace{1mm}
\noindent
\textbf{Distortion in exposure of producers / brands:}
As stated in Section~\ref{Sec: ExpBias}, the exposure of a brand / producer is defined as the sum of exposures of all items of that brand.  
We will now analyze the exposures received by various brands. 

Figure~\ref{Fig: AbsBrandBattery} shows the exposure values of the top-10 brands (as per the number of items they have in our battery dataset) in both the sponsored and organic RINs. 
A few of the top brands (e.g., Generic, Power one, Panasonic) see a significant drop in exposure in the Sponsored RIN, as compared to that in the Organic RIN.
The drop for the first two brands is more than 50\%, while for Panasonic it is nearly 35\%.
In contrast, the exposure of Amazon PL brands (highlighted within a rectangle) (+160\%) and Duracell (+43\%) see a significant increase in the sponsored RIN. In fact, Amazon is the most exposed brand in the sponsored RIN (in contrast to Duracell in the organic RIN). 
Alarmingly, more than 75\% of all {\it 3P} brands were found to be significantly under-exposed in the sponsored RIN. 

Qualitatively similar observations are found in case of backpack category also as shown in Figure~\ref{Fig: AbsBrandBackpack}.
Some of the top brands (e.g., Skybags, American Tourister, Wildcraft etc.) see a significant drop in exposure in the sponsored RIN as compared to that in the organic RIN.
The drop for American Tourister is around 38\% while that for the other two brands is more than 70\%.
In contrast, the exposure of Amazon PL brands and Cosmus have increased significantly in the sponsored RIN.
Similar to the observations for the battery category, more than 68\% of all {\it 3P} brands were found to be under-exposed in the sponsored RIN in the backpack category as well.

\subsection{Influence of the sensitive attribute }\label{Sec: ResultNBR}

All the analyses reported in this section show that Amazon PL products get much higher exposure via sponsored recommendations, than what they get via organic recommendations. 
But we are yet to specifically investigate whether their private label status attribute actually influences their higher exposure in sponsored recommendations. 
To understand the effects of different attributes on the beneficial commodity (such as, in-degree, exposure etc.) received by items, we train a NBR model (see Section~\ref{sub:nbr-method}).

For our analysis, we consider the number of recommendations different items get (in-degree of the corresponding items) as the dependent variable (the beneficial commodity). 
Note that one can take different beneficial treatments (or their proxies) as the dependent variable. 
For the sake of this experiment, we consider it to be the promotion or in-degree of different products.

The independent attributes we considered are -- whether the product is a PL or 3P, \textit{quality of the product} as reflected by the average user rating, \textit{popularity} of the product as defined by the number of ratings, whether the seller of an item (who won the buy-box) is a Fulfilled by Amazon (FBA) or not, \textit{seller quality}  as reflected by the average user rating of the seller, and \textit{seller popularity} as reflected by the number of ratings a seller has got. 
The objective is to predict the in-degree of a product given the above attributes, and hence to see which attribute most influences the in-degree (promotion) of items.

\begin{table}[t]
	\noindent
	\small
	\centering
	\begin{tabular}{ |p{2.2 cm}|p{1.25 cm}|p{1.25 cm}||p{1.25 cm}|p{1.25 cm}|}
		\hline
		\textbf{Attributes} & \multicolumn{2}{c||}{\textbf{Backpack}}  & \multicolumn{2}{c|}{\textbf{Battery}}\\
		\hline
		& Organic & Sponsored & Organic & Sponsored \\ 
		\hline		
		Intercept & 2.317** & 1.296** & 1.467** & 1.348** \\
		\hline 
		Seller fulfillment & 0.262** & 1.097** & 0.270** & 0.791** \\
		\hline
		\textbf{PL membership} & \textbf{-0.183}* & \textbf{0.742**} & \textbf{0.961**} & \textbf{2.738**} \\
		\hline
		Product quality & 0.105** & 0.233** & 0.255** & 0.087** \\
		\hline
		Product popularity & 0.001** & 0.001** & 0.0** & 0.003** \\
		\hline
		Seller quality & 0.120** & 0.161** & 0.119** & 0.248** \\
		\hline
	\end{tabular}	
	\caption{{\bf Attributes along with their estimates as per the NBR model fitted on the dataset inferred from sponsored and organic RIN. Seller fulfillment and PL membership attributes have the strongest influence on the in-degree of items. * indicates $p <0.05$, and ** indicates $p<0.01$.}}
	\label{Tab: NBRAmazon}
	\vspace{-9 mm}
\end{table}

\noindent
\textbf{Observations:} 
The attributes and their estimated correlation values (as output by NBR) are listed in Table~\ref{Tab: NBRAmazon}. 
The seller being an FBA (fulfillment attribute) has the most significant effect in determining the in-degree of a product in both RINs of the backpack category; however for battery category, the most significant attribute is the PL membership of the product. PL membership is overall the most significant attribute of all in estimating the in-degree of items. 

The estimates of PL membership feature tell an interesting story on the two RINs of the backpack category. 
Being a PL product increases the chance of being recommended a higher number of times in the sponsored RIN (estimate $0.741$), while it has a {\it negative} estimate on the organic RIN ($-0.183$). This indicates that in practice, customers do {\it not} tend to visit the PL backpacks more while surfing for backpacks; however the same have been highly promoted in the sponsored RIN. Note that, among all features, the relative importance of PL attribute increases the most when we compare organic to sponsored RIN within the category.
Among other attributes, the quality of the product and that of the seller have positive estimates on the in-degree of products; whereas product popularity does not have any positive estimate on the in-degree. All the results were found to be statistically significant. However, in the sponsored RIN their relative importance in comparison to the aforementioned sensitive attributes is very low. 
From this analysis, it is clear that being a PL product specifically influences the number of sponsored recommendations received by an item.

\vspace{2mm}
\noindent
\textbf{Takeaways from the section:} We used multiple orthogonal methods and measures to estimate the relative bias toward Amazon PL products. 
Across all the analyses, we consistently observe that Amazon PLs are significantly more advantaged in the sponsored recommendations, as compared to the organic recommendations. 
Often these 
advantages for Amazon PL products come at the cost of disadvantages toward other 3P products.


\section{Concluding Discussion}

Our analysis using five different network-based biased measures suggests that Amazon PLs enjoy a significantly high promotion in the sponsored recommendations, compared to the organic (i.e., view similarity based) recommendations on Amazon.

Now, promotion of private label products is not illegal; in fact, many tech-giants regularly follow this practice. 
However, opaque, sponsored, private label recommendations can be abused by platforms to systematically evade competition in online marketplaces.
In turn, such policies may have long term economic consequences that affect the livelihood of millions of associated stakeholders.
We hope our findings would motivate researchers and practitioners to come up with methodologies and/or presentation strategies that would 
mitigate these biases.

\vspace{1 mm}
\noindent
\textbf{Toward mitigating exposure bias:} We believe that one of the best policies to circumvent this problem is to raise the `curtain' over the black-box algorithms. While Amazon has made some efforts in this direction~\cite{Finley2016Amazon}, we feel that this is still fragmented and more such efforts should be in place. 
Along with the algorithms, some level of transparency in the practices and policies that e-commerce marketplaces follow for placing these sponsored ads might be useful in the long run.
We understand that there are policy regulations and privacy issues that need to be considered; nevertheless, such issues can possibly be resolved through proper discussions and appropriate guidelines from the company's legal cell.



\vspace{1 mm} 
\noindent 
\textbf{Future directions:} An immediate next step in this line of work is to explore methods to mitigate the exposure bias without affecting the underlying notions of relatedness among items. Note that, sponsored results by definition will be different from their organic counterparts. However, we believe that reasonable arguments can be made for policies that allow sponsored exposure to deviate from organic exposure while thresholding on the extent of such deviation.
We plan to explore this direction in future. 




\begin{acks}
This research was supported in part by a European Research Council (ERC) Advanced Grant for the project ``Foundations for Fair Social Computing", funded under the European Union's Horizon 2020 Framework Programme (grant agreement no. 789373). A. Dash is supported by a fellowship from Tata Consultancy Services.
\end{acks}

\bibliographystyle{ACM-Reference-Format}
\bibliography{Main}

\balance

\end{document}